# The role of planetary formation and evolution in shaping the composition of exoplanetary atmospheres

Turrini D.[1,*], Nelson R. P.[2], Barbieri M.[3]

[1] Istituto di Astrofisica e Planetologia Spaziali INAF-IAPS, Via Fosso del Cavaliere 100, 00133, Rome, Italy
[2] School of Physics and Astronomy, Queen Mary University of London, Mile End Road, London, E1 4NS, U.K.
[3] Centro Interdipartimentale di Studi e Attività Spaziali CISAS, Università di Padova, Via Venezia 15, 35131, Padova, Italy

## Abstract

Over the last twenty years, the search for extrasolar planets has revealed the rich diversity of outcomes from the formation and evolution of planetary systems. In order to fully understand how these extrasolar planets came to be, however, the orbital and physical data we possess are not enough, and they need to be complemented with information about the composition of the exoplanets. Ground-based and space-based observations provided the first data on the atmospheric composition of a few extrasolar planets, but a larger and more detailed sample is required before we can fully take advantage of it. The primary goal of a dedicated space mission like the Exoplanet Characterization Observatory (EChO) proposal is to fill this gap and to expand the limited data we possess by performing a systematic survey of extrasolar planets. The full exploitation of the data that space-based and ground-based facilities will provide in the near future, however, requires knowledge about the sources and sinks of the chemical species and molecules that will be observed. Luckily, the study of the past history of the Solar System provides several indications about the effects of processes like migration, late accretion and secular impacts, and on the time they occur in the life of planetary systems. In this work we will review what is already known about the factors influencing the composition of planetary atmospheres, focusing on the case of gaseous giant planets, and what instead still need to be investigated.

## 1. Introduction

The primary objective of space missions devoted to the characterization of extrasolar planets, like the ESA M3 mission candidate Exoplanet Characterization Observatory (EChO), is to accurately measure transmission spectra for extrasolar planets during primary transits, and day-side spectra from secondary eclipses. An operating wavelength range in the near infrared and mid infrared (respectively NIR and MIR in the following) covers key absorption signatures from the molecular species $CH_4$, $H_2O$, $CO_2$ and CO, whose relative strengths scale with the abundances present in the planetary atmosphere. Chemical modelling shows that these abundances are particularly sensitive to the heavy-element content of the atmosphere (measured through the C/H and O/H ratios), and the C/O ratio (see Moses et al 2013, for example). Determination of the atmospheric elemental abundances from the measured spectra will provide important constraints on the formation, migration and enrichment history of the observed extrasolar planets.

* Email: diego.turrini@iaps.inaf.it
Phone: +39 06 4993 4414
Fax: +39 06 4993 4383

Spectroscopic observations of extrasolar planets using Hubble Space Telescope and Spitzer have already confirmed the existence of various elements and molecules such as sodium, water, methane, and carbon dioxide in the atmospheres of hot-Jupiters (e.g. Tinetti et al. 2007; Swain et al. 2009). Recent observations of the transiting hot-Jupiter Wasp-12b suggest an atmosphere abundant in CO and deficient in $H_2O$, consistent with an atmospheric C/O ratio > 1, in contrast to the solar value C/O=0.54 (Madhusudhan et al. 2011). Analysis of transmission and day-side spectra for the transiting 6.5 $M_{Earth}$ super-Earth GJ 1214b suggest a metal-rich atmosphere (e.g Bean et al. 2011), in agreement with the general expectation that low mass planets will be well-endowed with heavy elements. A similar conclusion has been reached for the hot-Neptune GJ 436b, whose day-side spectrum lacks a clear signature of $CH_4$ while displaying abundant CO and $CO_2$ (Madhusudhan & Seager 2011). The derived carbon chemistry mixing ratios are consistent with chemical models that assume a heavy element abundance enhanced above solar by a factor > 50 (Moses et al 2013). Although these and other data pertaining to extrasolar planetary atmospheres are tantalising, uncertainties originating in the relatively low signal to noise ratio, and low spectral resolution, mean that definitive conclusions concerning atmospheric abundances cannot be made. These data are not accurate enough to discriminate between different formation and migration scenarios for the observed planets. The spectral resolution and signal to noise ratio to be achieved by future space NIR/MIR missions similar to EChO will dramatically improve the situation and allow atmospheric compositions to be measured with unparalleled accuracy. Combining these data with estimates of planetary bulk compositions from accurate measurements of their radii and masses will allow degeneracies associated with planetary interior modelling to be broken (e.g. Adams et al. 2008), giving unique insight into the interior structure and elemental abundances of these alien worlds.

Understanding the link between the history of a planetary system and the atmospheric composition of its planets, however, is not only a problem of increasing and refining the sample of observational data. Also our comprehension of the formation and evolution of planetary systems needs to be expanded, and different open issues need to be addressed. The original view of the set of events and mechanisms involved in planetary formation, in fact, was derived from observations of the Solar System as it is today. The assumption derived from these observations was that planetary formation is a local, orderly process that produces regular, well-spaced and, above all, stable planetary systems and orbital configurations. However, with the discovery of increasing numbers of extrasolar planetary systems through ground- and space-based observations, it has become evident that planetary formation can result in a wide range of outcomes, most of them not obviously consistent with the picture derived from the observations of the Solar System.

The orbital structure of the majority of the discovered planetary systems seems to be strongly affected by planetary migration. This can arise through the exchange of angular momentum with the circumstellar disk in which the forming planets are embedded (see e.g. Papaloizou et al 2007, Chambers 2010 and references therein), and through the so-called “Jumping Jupiters” mechanism (Weidenschilling & Marzari 1996; Marzari & Weidenschilling 2002; Chatterjee et al. 2008), which invokes multiple planetary encounters with a chaotic exchange of angular momentum and energy between the bodies involved. Each of these migration mechanisms has different implications for the chemical make-up of the planetary atmosphere, as migration through a disk allows the planet to accrete from regions with varying chemical abundances. This is not true for hot Jupiters reaching their present orbital location through planetary scattering.

The growing body of evidence that dynamical and collisional processes, often chaotic and violent, can dramatically influence the evolution of young planetary systems gave rise to the idea that our Solar System could have undergone the same kind of evolution and represent a “lucky” case in which the end result was a stable and regular planetary system. As we will discuss shortly, different attempts at modelling have been performed on this regard, but in the context of NIR/MIR observations the underlying and important idea is that the processes shaping the formation and evolution of planetary systems are general. As a consequence, on one hand there are lessons that can be drawn from the Solar System and used to shed light on the link between the history of a planetary system and the atmospheric composition of its giant planets. On the other hand, the insight that will be provided by future NIR/MIR space missions similar to EChO will prove invaluable to improve our understanding of our own Solar System and of the processes that made it a favourable environment for the rise of life.

If we follow the description of the history of the Solar System by Coradini et al. (2011) and we generalize it, the life of planetary systems can be viewed as composed by three different phases (*circumstellar disk*, *protoplanetary disk*, fully formed *planetary system*), each characterized by different physical processes and different durations. This schematic view of the evolution of planetary systems is summarized in Fig. 1 (adapted and expanded from Coradini et al. 2011), where we report the main events that can take place across the different phases. Giant planets must form during the circumstellar disk phase, since the gaseous component of the disk is required to provide the material for both the massive envelopes of Jupiter-like planets and the limited ones of Neptune-like planets. Given that the processes that give rise to super-Earths are plausibly the same producing the cores of giant planets, also super-Earths can in principle complete their formation already during the circumstellar disk phase. Based on the case of the Solar System, terrestrial planets should instead complete their accretion after the dispersal of the gaseous component of the circumstellar disks.

In the following we will not discuss the details of the mechanisms governing planetary formation: interested readers can find updated reviews on the formation of the terrestrial planets in Righter & O'Brien (2011) and Morbidelli et al. (2012), on the formation of the giant planets in D'Angelo et al. (2011), and on the subject of planetary migration in Papaloizou et al. (2007) and Chambers (2009). We will instead focus the discussion on the processes and events affecting the atmospheric composition of the giant planets, as they are the major players in shaping the evolution of planetary systems and they constitute the largest fraction of the targets that will be observed by future NIR/MIR instruments .

## 2. Planetary formation and composition

The EChO mission proposal aimed to target super-Earths, Neptune-like and Jupiter-like exoplanets on relatively short period orbits. These broad classes of planets are all expected to have very different formation and migration histories that will be imprinted on their atmospheric and bulk chemical signatures. Within each of these planet taxonomic classes, the stochastic nature of planetary formation will be reflected in significant variations in the measured abundances, providing important information about the diverse formation and migration pathways experienced by planets that are members of the same broad class. Reconstructing formation histories from spectral measurements presents a challenging inversion problem, but can nonetheless provide very useful constraints, as we detail below.

Formation processes and migration influence a planet’s composition in numerous ways. For example, we clearly expect gas giant planet formation via gravitational instability to result in very different bulk compositions and atmospheric abundances compared with planets that form through core accretion. In general one may expect planets formed via the former process to reflect the bulk composition of the nascent circumstellar disk, whereas planets formed through core accretion can display a range of abundance ratios that depend on the relative accretion rates for planetesimals and gas (see Fig. 2). Gas giant planets are expected to have atmospheric compositions very different from the presumably heavy element-rich atmospheres of super-Earths, if the examples of Uranus and Neptune in our Solar System provide a useful guide. Very little research has been done on this important question, mainly because the large uncertainties in current measurements of elemental abundances provide little in the way of discrimination between different models and scenarios. Future NIR/MIR missions similar to EChO will change this, stimulating in-depth analyses of the link between formation, migration, and post-formation enrichment. In the absence of existing detailed model results, we outline a number of different simplified formation and migration scenarios to illustrate how diverse atmospheric elemental abundances can arise.

Formation of gas giants through gravitational instability during the earliest phases of a circumstellar disk evolution will lead initially to atmospheric abundances that are essentially the same as the central star's (see Fig. 2). Recent studies show that rapid inward migration of planets formed in this way occurs on time scales ~ $10^3$ yr (e.g. Baruteau et al 2011, Zhu et al 2012), too short for significant dust growth or planetesimal formation to arise between formation and significant migration occurring. As such, migration and accompanying gas accretion should maintain the initial planetary abundances. As we describe later, post-formation enrichment may occur through bombardment from neighbouring planetesimals or long-period star-grazing comets, but this enrichment will occur in an atmosphere with abundances that are essentially equal to the stellar values and will take place at a time in which the population of planetesimals is reduced respect to its initial value.

In its simplest form, the core accretion model of planet formation begins with the growth and settling of dust grains, followed by the formation of planetesimals that accrete to form a planetary core. Growth of the core to a mass in excess of a few Earth masses allows settling of a significant gaseous atmosphere from the surrounding nebula. Halting growth at this point results in a super-Earth or Neptune-like planet. Continued growth through accretion of planetesimals and gas can lead to runaway gas accretion, forming a Jupiter-like gas giant. A key issue for determining the atmospheric abundances of a forming planet is the presence of ice-lines at various distances from the central star, where volatiles such as water, carbon dioxide and carbon monoxide freeze-out onto grains and are incorporated into planetesimals. Considering a typical protoplanetary disk orbiting a solar-type star, Fig. 3 shows that a $H_2O$ ice-line is expected at ~2 au, a $CO_2$ ice-line at ~10 au, and a CO ice-line at ~ 40 au. The atmospheric abundances of a planet during formation will therefore depend on where it forms and the ratio of gas to solids accreted at late times (see Fig. 3). We recall that the EChO mission proposal focused on the study planets with relatively short orbital periods, and these must have undergone large scale migration during their evolution: the inner regions of circumstellar disks contain too little solid material for *in situ* formation of planetary systems similar to those discovered by the Kepler mission (e.g. Kepler 11, Lissauer et al. 2011) and radial velocity surveys (e.g. Gliese 581, Mayor et al. 2009). The final C/O ratio of the planet will therefore depend on how it accretes as it migrates through the disk.

For the purpose of illustration, we now consider a number of highly simplified planetary accretion and migration scenarios and their influence on the atmospheric C/O ratio, which a space mission similar to EChO would measure for tens of giant planets (see EChO Science Study Team 2013 and, specifically, the "Origins" observational tier of EChO). The circumstellar disk is assumed to be of solar abundance, giving rise to an overall C/O ratio ~ 0.54, as shown in Fig. 3. Interior to the $H_2O$ ice-line, carbon- and silicate-rich grains are condensed, leading to an increase of the gas-phase C/O ~ 0.6 (due to the slight overabundance of oxygen relative to carbon in these refractory species). Water condenses between the $H_2O$ and $CO_2$ ice-lines, increasing the gas-phase C/O ~ 0.85 through removal of oxygen into the frozen-out $H_2O$, and decreasing the solid phase C/O ~ 0.26. Between the $CO_2$ and CO ice-lines the $CO_2$ freezes out, increasing the gas phase C/O ~ 1 and moderately increasing the solid phase C/O ratio ~ 0.31. Various formation scenarios for gas giant planets are now assumed, based on Figs. 2 and 3, and their implications for the atmospheric C/O value are calculated.

<u>Scenario 1</u>: A solid core forms at 5 au, undergoes type I migration inward, and only starts to accrete gas once it has moved interior to the $H_2O$ ice-line. Gas accretion is not accompanied by any accretion of solids, leading to an atmospheric C/O ratio ~ 0.6. Such a scenario may apply when an earlier episode of planetary formation has occurred in the disc interior to 5 au, depleting the planetesimals that were orbiting there.

<u>Scenario 2</u>: This is identical to scenario 1, except that gas accretion interior to the $H_2O$ ice-line is accompanied by accretion of solids, such that the C and O abundances of the accreted material equals the solar value. The atmospheric C/O ratio is ~ 0.54. Here the inner disc would not have experienced an earlier episode of planet formation, so planetesimals are available for accretion by the migrating giant as it migrates into this region and accretes gas.

<u>Scenario 3</u>: A solid core forms at 5 au and accretes gas exterior to the $H_2O$ ice-line while migrating inward, without accreting any solids. Before crossing the $H_2O$ ice-line the planet opens a deep gap that prevents further accretion. The planet continues to migrate inward. The final C/O ratio is ~ 0.86. In this scenario, the core of the giant planet must grow quickly as it forms at 5 au, allowing it to reach the critical core mass prior to migrating. Rapid gas accretion can then occur prior to gap formation and migration through the $H_2O$ ice-line.

<u>Scenario 4</u>: A solid core forms at 15 au and accretes gas but no further solids from beyond the $CO_2$ ice-line before forming a deep gap that prevents further gas accretion. The planet migrates inward to form a hot Jupiter and the final atmospheric C/O ratio is ~ 1. In this scenario, the solid core of the planet needs to form and initiate gas accretion at quite large radius from the star. This may occur if planetary cores can migrate outward because of the strong influence of corotation torques, as considered by Paardekooper et al (2011) and Hellary & Nelson (2012). Rapid gas accretion leading to a gap-opening planet is likely to require the formation of a massive core out at 15 au.

<u>Scenario 5</u>: A solid core forms at 15 au, migrates inward, and accretes equal amounts of gas from the region outside the $CO_2$ ice-line, the region between the $H_2O$ and $CO_2$ ice-lines, and the region interior the $H_2O$ ice-line. No solids are accreted. The final atmospheric C/O ratio is ~ 0.77. This scenario can arise if outward migration of planetary cores occurs, as described for scenario 4, but gas accretion occurs more gradually as the planet migrates inward. Here the core mass is likely to be smaller than in scenario 4, and/or the opacity of the gaseous envelope will be higher, leading to slower gas accretion.

<u>Scenario 6</u>: A solid core forms at 15 au, migrates inward and accretes gas as in scenario 5, but also accretes 10% of its atmosphere in the form of solids as it migrates. The final atmospheric C/O ratio is ~ 0.73. The scenario arises when the conditions for scenario 5 are satisfied, and the inner disc regions retain a population of planetesimals that can be accreted by an incoming, migrating giant planet.

All of the above scenarios assume gas-dominated accretion and lead to solar or super-solar atmosphere C/O ratios because of the tendency of O-rich compounds to freeze-out at higher temperatures. Sub-solar values of the C/O ratio can be obtained through substantial accretion of silicate-rich planetesimals as the planet migrates interior to the $H_2O$ ice-line. These examples simply serve to illustrate that a variety of formation, migration and accretion scenarios can lead to a broad distribution of C/O ratios. The final C/O value correlates with where and how the planet forms and migrates in a predictable manner, but this final value is not unique for all scenarios. Detailed predictions of the expected diversity of C/O ratios in planetary atmospheres require planetary formation models to be computed that account for the evolving chemistry of the protoplanetary disk and the chemical abundances of the accreted material.

Finally, we note that gas disk-driven migration is only one plausible mechanism by which planets may migrate. As discussed previously, the large eccentricities (and obliquities) of the extrasolar planet population suggest that planet-planet gravitational scattering ("Jumping Jupiters") may be important (e.g. Weideschilling & Marzari 1996; Marzari & Weidenschilling 2002; Chatterjee et al 2008), and this is likely to occur toward the end of the gas disk lifetime when its ability to damp orbital eccentricities is diminished. When combined with tidal interaction with the central star, planet-planet scattering onto highly eccentric orbits can form short-period planets that have not migrated toward the central star while accreting from the circumstellar disk. These planets are likely to show chemical signatures that reflect this alternative formation history, being composed of higher volatile fractions if they form exterior to the $H_2O$ ice-line.

## 3. Post-formation evolution, late accretion and protoplanetary disks

Immediately after (~$10^5$ years) they form, giant planets trigger a phase of intense remixing of solid material in the protoplanetary disk in which they are embedded (Safronov 1969; Weidenschilling 1975; Weidenschilling et al. 2001; Turrini et al. 2011, 2012; Coradini et al. 2011), which manifest as a bombardment on the other planetary bodies populating the forming system. In the Solar System this event has been named the Jovian Early Bombardment (Turrini et al. 2011, 2012; Coradini et al. 2011; Turrini 2013; Turrini & Svetsov 2014) as Jupiter was likely the first planet to form (Safronov 1969; Coradini et al. 2011). As a new phase of remixing and bombardment, likely of decreasing intensity, will be triggered by the formation of each giant planet in a planetary system hosting more than one, a more general name for this class of events is the Primordial Heavy Bombardments (Coradini et al. 2011). The duration of the phase of bombardment and remixing triggered by the formation of Jupiter in the Solar System was estimated to be of about 0.5-1 Ma (Weidenschilling 1975; Turrini et al. 2011, 2012). The bombardment is caused by the interplay between the gravitational scattering of planetesimals near-by the newly formed giant planet and the appearance of the orbital resonances in regions farther away (Safronov 1969; Weidenschilling 1975; Weidenschilling et al. 2001; Turrini et al. 2011, 2012; Turrini 2013; Turrini & Svetsov 2014).

From the point of view of future NIR/MIR observations by space missions similar to EChO, the most important effect of this class of events is the reshuffling of the solid material present in the circumstellar disks: volatile-rich objects from beyond the Snow Line are injected into the inner, volatile-depleted regions of the disk while rocky, metal-rich bodies are transferred from the latter to the former. The net effect is a change in the rock-ice, metal-ice and likely of the $H_2O$-$CO_2$ ice ratios in the different regions of the circumstellar disks, as can be seen by comparing Figs. 3 and 4. During this phase of remixing, the orbital regions of the giant planets are crossed by these fluxes of planetesimals and a fraction of the migrating material is captured by the giant planets themselves, as already pointed out by Weidenschilling (1975) for the case of Jupiter, thus enriching their atmospheric composition in high-Z elements.

To illustrate the effects of these events and the implications for the interpretation of NIR/MIR observations, we will take advantage of the following toy model. We consider a Jupiter-sized giant planet disk embedded into a disk of massless particles representing the circumstellar disk. Dynamical friction between the bodies populating the circumstellar disk and the effects of gas drag are ignored for simplicity. The giant planet (see Fig. 2) is initially on a planar ($i = 0°$) and circular ($e = 0$) orbit with the semimajor axis of Jupiter ($a = 5.2$ au). The massless particles are distributed between 1 and 10 au, leaving empty the region between 4.7 and 5.7 au to simulate the gap created by the formation of the giant planet. The massless particles initially have eccentricities randomly distributed between 0 and 0.1 (Weidenschilling 2008) and inclinations randomly distributed between 0° and 1.7° (Turrini et al. 2011). The $H_2O$ ice-line is assumed to be at 4.0 AU and the massless particles are divided into compositional classes according to their semimajor axes. Using the Solar System as a template, bodies inside 3 AU are considered composed of a mixture of rocks and metals analogous to ordinary chondrites. Bodies in the region between 3.0 AU and 4.0 AU are assumed transitional bodies depleted in metallic iron (most iron is in oxydized form) and enriched in carbon and water similarly to the carbonaceous chondrites. Finally, bodies from beyond 4.0 AU are assumed to be volatile-rich bodies similar to the comets.

We consider three scenarios:

- the giant planet not migrating (i.e. standing on its initial orbit, hereafter labelled as the *no migration* case),
- the giant planet migrating to a semimajor axis of 0.7 AU with an e-folding time of $5\text{x}10^3$ years (i.e. 99.4% of the migration is completed in $2.5\text{x}10^4$ years, hereafter labelled as the *fast migration* case), and
- the giant planet migrating to a semimajor axis of 0.7 AU with an e-folding time of $3\text{x}10^4$ years (i.e. 99.4% of the migration is completed in $1.5\text{x}10^5$ years, hereafter labelled as the *slow migration* case).

Migration, when occurring, always starts after $10^4$ years from the beginning of the simulations. The migration scheme is implemented following Hahn & Malhotra (2005). As shown in Fig. 5, about 33% to 50% of the impacting particles are accreted by the giant planet extremely quickly in the first $10^4$ years. Then, if the giant planet migrates, the late accretion phase slows down significantly for the first 2 e-folding times (i.e. the faster part of the migration, while the giant planet complete about 86% of its displacement). Accretion starts again across the next 3 e-folding times (i.e. the slowest part of the migration, while the giant planet complete about 13% of its displacement). If the giant planet does not migrate, the late accretion extends over about $5\text{x}10^5$ years.

Accretion in the no migration case is the most efficient, with the giant planet capturing about 4.8% of the solid material in the circumstellar disk. The slow migration is more efficient accretion-wise than the fast migration case, with 3.8% vs 3.3% of the solid material of the disk captured by the giant planet. Assuming the disk is analogous to the Minimum Mass Solar Nebula and has a surface density profile governed by the relationship $\sigma = 2700\ r^{-3/2}$ g cm$^{-2}$ (Coradini et al. 1981) where $r$ is the orbital distance in au, the mass of the disk we considered would be about 27.5 $M_\oplus$ (where the symbol $M_\oplus$ indicates the mass of the Earth). Assuming that the giant planet had a core of 5 $M_\oplus$, this leaves a total of 22.5 $M_\oplus$ in the massless particles. The accreted masses would then translate in 1.1 $M_\oplus$ (no migration scenario), 0.86 $M_\oplus$ (slow migration scenario) and 0.75 $M_\oplus$ (fast migration scenario).

As the dynamical and physical model underlying this toy model is quite simplistic, these numbers should be regarded just as more detailed back-of-the-envelope calculations. Yet, they provide a first indication of the effects of the post-formation accretion. While the overall accretion efficiency varies between the three scenarios, Fig. 6 shows that the relative importance of the different source regions in the circumstellar disk varies little. This means that a newly formed giant planet can quickly accrete solid material from a vast feeding zone characterized by different compositions of the planetesimals. About 40% of the accreted bodies originate from the inner (1 - 4 AU) region of the protoplanetary disk.

If we use water and the most abundant elements (Si, C, N, S, Fe) as tracers of the composition of the accreted bodies, bodies originating between 1 - 3 AU would be composed (values derived from Jarosewich 1990) by Si (~19 wt%), Fe (~25 wt%), C (0.5 wt%), and S (1.8 wt%). Bodies coming from the 3 - 4 AU transition region would be composed (values derived from Jarosewich 1990) by Si (~14 wt%), Fe (~25 wt%), C (~2 wt%), S (1.8 wt%), and $H_2O$ (~10 wt%). Finally, half the mass of those bodies originating beyond the $H_2O$ ice-line (4 - 10 AU) will be assumed to have the same refractory-rich composition of carbonaceous chondrites, while the volatile-rich half of their mass will be mostly composed by $H_2O$ (~74 wt%), C-bearing molecules (~24 wt%), N-bearing molecules (~0.7%), and S-bearing molecules (~1.2 wt%), where the wt % ratios are estimated from the abundances reported by Mumma & Charnely (2011). Because of the condensation sequence of the different elements and chemical species in the protoplanetary disk (Lewis 2004), therefore, the post-formation accretion phase would bring to the giant planet high-Z materials with relative abundances of the different elements that are highly non-solar. As a consequence, the effects of this process should reflect into the C/O, S/O, N/O ratios and (possibly) in the content of silicates and metals in the atmospheres of the giant planets. Metals and silicates, however, had been observed by the Galileo spacecraft after the impact of the comet Shoemaker-Levy 9 on Jupiter only for a limited time (Taylor et al. 2004), which could imply that they are removed efficiently (i.e. in a matter of months) from the observable regions of the atmosphere of a giant planet in contrast to water, for example (see also Sect. “Secular contamination of planetary atmospheres”). As an example of the observable effects of late accretion during the Primordial Heavy Bombardment, in Table 1 we show the final mixing ratios for a Jupiter-like planet in the no migration scenario and the differences between the mixing ratios in the no migration, fast migration and slow migration scenarios and, as a reference, the differences between the no migration scenario and solar abundances (data from Irwin 2009) using the previously reported compositions of the planetesimals. The accreted mass is assumed to be distributed into a molecular shell of 5000 km (i.e. ~20% of the planetary mass, Guillot et al. 2004) of initial solar composition in a homogenous way, so that it the atmospheric composition is the same as that of the molecular shell. As a first approximation, we will ignore the effects of the chemistry induced by the temperature and pression profiles of the atmosphere (e.g. if C is in the form of $CO/CO_2$ it will reduce the amount of O available to form $H_2O$, while the same is not true if C is in the form of $CH_4$). As can be seen by comparing the values reported in Table 1 with the sensitivity of the three observational tiers of EChO (in order of increasing sensitivity: “Chemical Census”, “Origins” and “Rosetta Stones”; EChO Science Study Team 2013), a space mission similar to EChO would be capable of detecting the species associated to the elements reported in Table 1 already in its lowest sensitivity tier (i.e. the “Chemical Census”, EChO Science Study Team 2013). It is worth noting that, in principle, the sensitivity of the “Chemical Census” tier would be enough to detected enrichments respect to stellar abundances and to discriminate, for two otherwise similar planets, which one underwent migration and which one formed in situ: this indicates that a mission like EChO could provide us with a first look into the dynamical histories of hundreds of exoplanets during its operational lifetime. The intermediate sensitivity tier, the “Origins” tier (EChO Science Study Team 2013), could then allow to study in more detail the past orbital evolution of tens of exoplanets.

| Element/ Molecule | No Migration Mixing Ratios | No Migration vs Solar Abundances | No Migration vs Fast Migration | No Migration vs Slow Migration | Slow Migration vs Fast Migration |
|---|---|---|---|---|---|
| Fe | $7.7\times10^{-3}$ | $6.6\times10^{-3}$ | $7.5\times10^{-4}$ | $5.3\times10^{-4}$ | $2.2\times10^{-4}$ |
| Si | $2.2\times10^{-3}$ | $1.5\times10^{-4}$ | $4.8\times10^{-4}$ | $3.4\times10^{-4}$ | $1.3\times10^{-4}$ |
| C | $2.6\times10^{-3}$ | $4.7\times10^{-4}$ | $1.5\times10^{-4}$ | $9.8\times10^{-5}$ | $5.0\times10^{-5}$ |
| N | $6.6\times10^{-4}$ | $4.2\times10^{-5}$ | $1.3\times10^{-5}$ | $8.7\times10^{-6}$ | $4.5\times10^{-6}$ |
| S | $8.3\times10^{-4}$ | $4.9\times10^{-4}$ | $6.6\times10^{-5}$ | $4.6\times10^{-5}$ | $2.0\times10^{-5}$ |
| $H_2O$ | $9.0\times10^{-3}$ | $5.4\times10^{-3}$ | $1.2\times10^{-3}$ | $7.6\times10^{-4}$ | $4.0\times10^{-4}$ |

*Table 1: Mixing ratios of a Jupiter-like planet in the no migration scenario and differences in the mixing ratios between the no migration, fast migration and slow migration scenarios and between the no migration scenario and solar abundances (data from Irwin 2009), where the captured mass is homogeneously mixed into a 5000 km thick outer molecular shell of initial solar composition.*

As mentioned above, the reshuffling process started by the Primordial Heavy Bombardment has a duration of about 1 Ma. At its end, however, it transitions (in the classical scenario for the formation of the Solar System) into a longer phase of reshuffling and dynamical clearing where the planetary embryos in the protoplanetary disk scatter planetesimals inside the now-depleted locations of the orbital resonances with the giant planets. This phase has been studied, in the Solar System, to investigate the mass depletion of the asteroid belt (Wetherill 1992; Chambers & Wetherill 2001; Petit et al. 2001; O'Brien et al. 2007). During this phase, the population of planetesimals in the affected regions decays exponentially, decreasing by about two orders of magnitude in about 100 Ma (see e.g. O'Brien et al. 2007). Across these 100 Ma the giant planets continues to capture part of the solid material that is expelled by the resonances: in principle, this process could possibly be even more efficient in the case of giant planets on inner orbits (e.g. less than 1 AU) if they did not completely dispersed the planetary bodies populating the regions they crossed while migrating.

Limited data are currently available on the implications of the phase of dynamical clearing for the composition of the giant planets taking into account the role of planetary embryos. The best estimate to date is the one done by Guillot & Gladman (2000), who assessed the capture efficiency of the four giant planets during the 100 Ma following the formation. The work of Guillot & Gladman (2000) was based on a simplified model similar to the toy model we used to illustrate the effects of the Primordial Bombardment. The disk of massless particles extended from 4 AU to 35 AU and planetary embryos were not included in the simulations. The cumulative capture efficiency of the four giant planets was found to be about 4%, i.e. of the same level as the one found with our toy model for Jupiter alone. It would therefore appear that the combined perturbations of the giant planets, once they are all present in the planetary system, and the effects of concurrent accretion make late accretion very inefficient after the first few Ma (ejection from the planetary system is favoured). This is confirmed also by Guillot & Gladman (2000), who in a second set of simulations show that the accretion efficiency of Jupiter alone could rise up to 7-8% in a disk extending between 4 AU and 13 AU, so twice as much as that of the four giant planets cumulatively.

Extensive migration of the giant planets, both due to the interaction with the disk or to planet-planet scattering (i.e. the Jumping Jupiter mechanism, Weidenschilling & Marzari 1996; Marzari & Weidenschilling 2002; Chatterjee et al. 2008) can supply an alternate evolutionary path to the previously described picture derived from the Solar System. The dynamical effects of the migrating planet on the protoplanetary disk would replace the slow erosion due to the interplay between planetary embryos and orbital resonances with the giant planets, but the end result would be analogous: the reshuffling of material from different regions of the protoplanetary disk and the depletion of the population of planetesimals with a consequence capture of part of the removed population by the giant planets. As our toy model shows for a very simple configuration of the planetary system, the migrating giant planet would still capture material from a wide range of orbital distances. Another, more complex example is constituted by the so-called “Grand Tack” scenario (Walsh et al. 2011, 2012), where the four giant planets of the Solar System are suggested to have migrated inward, then outward and to get locked in the resonant configuration they were suggested to be before the Late Heavy Bombardment (Levison et al. 2011). During their extensive migration, the giant planets would scatter and redistributed the primordial planetesimals in the Solar System and likely capture a fraction of them. While it has been argued that this scenario has a low probability of reproducing the actual configuration of the Solar System (D’Angelo & Marzari 2012), the richness of orbital configurations of the known extrasolar planets can imply that several of the other planetary systems differing from our Solar System can be the outcome of the “failed” cases of this kind of evolutionary path. The results of Guillot & Gladman (2000) on the concurrent accretion and of our toy model for the accretion efficiency of migrating planets suggest, however, that in such a scenario the fraction of captured material (i.e. the late accretion) would be low.

A Jumping Jupiters evolution can also take place at a later time as has been hypothesized in the case of the Solar System by the so-called Nice Model (Gomes et al. 2005; Tsiganis et al. 2005; Morbidelli et al. 2005). The Nice Model is a Jumping Jupiter scenario formulated to link the event known as the Late Heavy Bombardment (Tera et al. 1974), assumed to have occurred about 600-800 Ma after the formation of the Solar System (see Fig. 1) to a migration event involving all the giant planets. In the Nice Model, the giant planets of the Solar System are postulated to have been initially located on a more compact orbital configuration than their present one and to interact with a massive primordial trans-neptunian region. The gravitational perturbations among the giant planets are initially mitigated by the trans-neptunian disk, whose population in turn is eroded. Once the trans-neptunian disk becomes unable to mitigate the effects of the interactions among the giant planets, the orbits of the latter become excited and a series of close encounters takes place. The end result of the Jumping Jupiters mechanism in the Nice Model is a small inward migration of Jupiter and marked outward migration of Saturn, Uranus and Neptune (Tsiganis et al. 2005; Levison et al. 2011). Due to the late time at which this migration and the associated bombardment take place and the depletion previously occurred in the population of planetesimals, however, the amount of solid material captured by the giant planets across the Late Heavy Bombardment would be limited. Matters et al. (2009) estimated that the accreted material would amount to 0.15 $M_\oplus$ for Jupiter, 0.08 $M_\oplus$ for Saturn and ~0.05 $M_\oplus$ for Uranus and Neptune. In the case of Jupiter, the material accreted during such a late event would be about an order of magnitude lower than the one accreted immediately after its formation.

These results collectively seem to indicate that the late accretion phase can in principle extend over a long temporal interval (500 Ma – 1 Ga) but that the magnitude of its effects decreases quickly with time, so that the main role is played by the first few Ma after the formation of a giant planet. Late accretion then turns into a slow, secular contamination process whose temporary effects, however, can have implications for future NIR/MIR observations and their interpretation, as will be illustrated in the next section.

## *4. Secular contamination of planetary atmospheres*

Once they complete the most active and violent phases of their evolution, planetary systems enter a stationary phase governed by secular processes. During this phase, the main processes affecting the atmospheric composition of the planets are impacts, atmospheric chemistry and, in the case of short-period planets, the stellar radiation and wind. Impacts, in particular, allow the transfer of material between different planetary bodies and between different orbital regions, continuing the remixing process that acted across the early phases of the evolution of planetary system but at a much lower rate. An example of this process in the Solar System is represented by the flux of comets impacting Jupiter, the most famous (and studied) of which is the impact of comet Shoemaker-Levy 9 (SL9 in the following) in 1994. Across the last 17 years the giant planet has been hit by five impactors: SL9 itself in 1994 (~5 km in diameter), then a sub-km impactor in 2009 (~10 m in diameter), two in 2010 (~500 m in diameter the first and undetermined the second) and one in 2012 (< 10 m in diameter). A sixth impact of a meteoroid was observed by Voyager 1 in 1979 (estimated mass of 11 kg, i.e. ~10 cm in diameter, Cook & Duxbury 1981). It must be stressed that these impact rates do not represent the real flux on Jupiter, but they are more likely a reflection of the observational coverage of the giant planet.

Recent results from the Herschel mission (Cavalié et al. 2013) indicate that the spatially-resolved distribution of stratospheric water for Jupiter cannot be explained by local processes or a steady state flux of interplanetary dust particles, and are instead a reflection of the impact of comet SL9. In particular, about 95% of the stratospheric water content of the giant planet as been reported to be due to this cometary impact. Similar but non-spatially resolved results were previously obtained for both water, CO and $CO_2$, supporting the case for an external source for these molecules as the thermal and pressure profiles in the atmosphere of Jupiter create a transport barrier between troposphere and stratosphere (see Cavalié et al. 2013 and reference therein), where water is expected to condense. The mixing ratio of stratospheric water modelled from the measurements of Herschel is $1.7 \times 10^{-8}$ (Cavalié et al. 2013). The coupling between the long persistence of water in the stratosphere to Jupiter and its limited mixing ratio is of particular importance, because the detection of such small mixing ratios requires NIR/MIR observations over long timescales, comparable to the persistence of water itself.

To link the atmospheric composition of exoplanets to their formation it is therefore mandatory to be able to discriminate the (plausibly) transient contribution in high-Z elements due to external sources from the constant one due to the bulk composition of the planet. To get a zero-order estimate of the implications of secular, cometary impacts for the observations we can use a simple toy model. We consider the planet HD 189733 b as our test case. For this planet atmospheric water has been detected with a model mixing ratio of $5 \times 10^{-4}$ (Tinetti et al. 2007), which is well inside the observational capabilities of EChO of the “Chemical Census” tier described in the EChO Yellow Book (EChO Science Study Team, 2013). Orbital and physical parameters for the host star and the planet were obtained from the Extrasolar Planets Encyclopaedia (www.exoplanet.eu). If we consider an atmospheric shell analogous in size to the Jovian stratosphere, i.e. about 300 km and with a average density of $2 \times 10^{-8}$ g cm$^{-3}$ (from the geometric mean of the extreme values reported by Young et al. 2005 for the Jovian stratosphere), the water content of this shell would be about $2.5 \times 10^{17}$ g.

As possible impactors, we consider a population of star-grazing exocomets modelled after the 1265 Sun-grazing comets observed by SOHO since 1996: their orbital parameters were obtained from the JPL Small Bodies Database Search Engine (http://ssd.jpl.nasa.gov/sbdb_query.cgi). Sun-grazing comets observed by SOHO have high orbital inclination values (i.e. ~140°) therefore our preliminary estimate gives 1 impact on our test planet every 200 years. Assuming an ecliptic population of star-grazing exocomets, our preliminary estimate gives the larger flux of 1 impact every 20 years. The 100 comets for which we have estimates of the diameter range in size between 0.5 km to 60 km. If we consider the whole sample, the average diameter is about 5 km, i.e. the estimated size of comet Shoemaker-Levy 9, while if we ignore comets larger than 10 km, to compensate for the observational bias favouring larger objects respect to smaller ones, the average diameter is about 3 km. We will consider this value as the reference one.

Assuming a density of 1 g cm$^{-3}$, the mass of these bodies would range between $5 \times 10^{14}$ g (1 km) and $6.5 \times 10^{16}$ g (5 km), with our reference case (3 km) being $1.4 \times 10^{16}$ g. From Mumma & Charnley (2011), we can assume an average water content of 74% of their mass. Therefore, each of these impactors would bring, on average, $10^{16}$ g. This implies that it would take the cumulative water budget of about 25 cometary impacts to reproduce the water content of the atmospheric shell we considered. Equivalently, the water content delivered by cometary impactors should be able to survive in said shell between $5 \times 10^{2}$ years (1 impact every 20 years) and $5 \times 10^{3}$ years (1 impact every 200 years). These values obviously assume that all the cometary water is released in the stratosphere of the exoplanet, which is not necessarily the case. Note that also this toy model should be regarded just as a more sophisticated back-of-the-envelope calculation, as the uncertainties on both the observational constrains and the assumptions can plausibly affect the order of magnitude of the values here considered.

As an example of the uncertainties affecting these estimates, Tinetti et al. (2007) report that changing the mixing ratio of water in the atmosphere of HD 189733 b by plus or minus one order of magnitude does not significantly affect the fit to the observational data. This uncertainty on the mixing ratio has significant implications for the time required to accumulate the expected water content. In the most favourable case (i.e. a mixing ration of $5 \times 10^{-5}$), which would still be detectable in the "Chemical Census" tier described in the EChO Yellow Book (EChO Science Study Team 2013), it would take only a couple of impacts to provide the required water budget or, alternatively, water would need to survive in the atmospheric shell only about 50 years (consider that water from the Shoemaker-Levy 9 impact resided in the Jovian stratosphere for at least 14 years at the time of the observations used by Cavalié et al. 2013). In the most unfavourable case, cometary water would need about $5 \times 10^{4}$ years to accumulate.

From an observational perspective, the effects of the secular contamination can be expressed in terms of the mixing ratio that cometary impactors would produce. From the previous calculations, we can derive the following empirical relationship for the mixing ratio produced by a cometary impactor in the stratosphere of an exogiant planet:

$$m_r = 2.98 \times 10^{-6} \chi \left( \frac{D_{comet}}{1\,km} \right) \left( \frac{\rho_{atmosph.}}{2 \times 10^{-5}\,kg\,m^{-3}} \right)^{-1} \left( \frac{\Delta R}{100\,km} \right)^{-1}$$

where $m_r$ is the mixing ratio of atmospheric water contributed by the impactor, $\chi$ is the fraction of the impactor dissolved in the considered atmospheric shell (i.e. the effective amount of water delivered in the observable region) and *ΔR* is the thickness of the atmospheric shell considered. This relationship can be used to obtain a rough estimate of the effects of cometary impactors for what it concerns water, but similar relationships can be derived for other high-Z elements and populations of impactors.

## 5. Planetary processes affecting atmospheric composition after cometary or asteroidal impacts

As we mentioned above, the study of impact of small bodies with giant planets took a great advantage of the impact of SL9 on Jupiter in 1994. In the following years other small bodies impacted with Jupiter, confirming the idea that these events are not so rare in the history of the Solar System. The analysis of the SL9 impact data and the application of the theory of atmospheric entry of meteor on Earth made it possible to understand the physical behaviour of the Jovian atmosphere after the impact. Each fragment of SL9 reached a different depth inside the planet, before exploding and releasing all his energy to the surrounding atmosphere. The maximum depth reached depend only on the physical properties of the impactor as: size, density, velocity and angle of entry. The larger impactors, about 3 km in size and mainly composed of water ice, reached a maximum depths of about 500 km (approximatively 1 Kbar). At the maximum depth the fragments exploded and liberated all their energy. In a time-scale of few second the hot gases generated by the explosion have risen to the higher parts of the Jovian atmosphere characterized by lower opacities, where they reached a luminosity of the order of $10^{25}$erg/s in the near-infrared.

The sudden release of this large amount of energy created a shock wave that propagate at high velocity trough the atmosphere, with the shocked material reaching very high temperatures and pressures. In this condition it is possible to form new chemical species and transport them over great distances in the atmosphere. In the case of SL9 impact the most surprising report was that of $S_2$ at impact site of G fragment. Newly detected or enhanced molecular species resulting from impacts included: CS, $CS_2$, OCS, $H_2S$, $SO_2$, HCN, CO, $H_2O$, $NH_3$, $C_2H_4$, $PH_3$. All new and enhanced species were detected in Jupiter's stratosphere. A general rule of shocked chemistry is that CO forms until either C or O is exhausted. If O > C, the other products are oxidized, and excess O goes to $H_2O$. If C > O, the other products are reduced, and excess C goes to HCN, $C_2H_2$, and a wide variety of more complicated organics. The dark ejecta debris is probably composed in part of carbonaceous particles generated by the shocks. The observed O/S ratio was reasonably consistent with cometary abundances, but the O/N ratio was much larger, suggesting that another N species was formed (presumably $N_2$) but probably remained undetected. Convective motions in the upper part of the atmosphere, and atmospheric circulations then provided a rapid mixing of the newly formed species over short time scales (few days) in the planetary atmosphere, but of the observed species OCS, $H_2S$ and $NH_3$ were found to be transient on a time scale of months. $CS_2$ and CO were observed to weaken over time and to be still present in the atmosphere 1 year after the impact. In contrast, spectral signatures of HCN and CS remained strong also 1 year after the impact, indicating that their abundances remained constant or even increased with time.

Despite the great changes induced by impact of SL9 on Jupiter it will be difficult to observe with the same details a similar event on an extrasolar planet and to infer the nature of changes in its atmospheric chemical composition as induced by a single impact. However, assuming that the phases underwent by the Solar System in the earlies epochs of its life are common to all planetary systems in the Galaxy, there is a series of event that could alter in a secular way the chemical composition of an extrasolar planet. As we also mentioned before, the migration of massive bodies like Jupiter is associated to large perturbations on smaller bodies presents in the extrasolar analogues of the asteroidal belt or of the Oort clouds, displacing them from their original orbits and injecting in more eccentric orbits that could intersect also the central star. During the migration, the planet could be hit many times by these small bodies: even if the exogenous material delivered in the exoplanetary atmosphere can be limited, especially during late events as illustrated in the previous sections, the changes induced by the energy input associated to the impacts can be relevant.

If the planet has a radiative zone in the upper part of its atmosphere, as suggested by Guillot et al. (1995) in the case of Jupiter, it can extend from 500 to 1000 km (see Fig. 7). This radiative zone act as a barrier between the planet interior and the upper atmosphere, impeding any exchange of material between the two convective regions, preventing in this way a full mixing of the internal material with the external zones. The same situation happens in the main sequence stars of spectral type F where a radiative zone separate the external atmosphere from the inner region where convection is active and nuclear burning products are well mixed. In the case of an impactor with size of about 10 km on Jupiter, with velocity and impact angle similar to those of SL9, the maximum depth reached by the body could be larger than 1000 km, well below the radiative zone. In this case the fireball and the shock wave created by the impact will expand in a region with different chemical composition (more He rich) respect to the upper atmosphere. At the same time the snowball effect created by the shock wave will be able to bring He rich material above the radiative zone, hence altering the atmospheric chemical composition of the planet. The material below the radiative zone will be brought to the surface, while the upper atmospheric material is depleted in heavier elements and chemical species.

These events have a strong similarity with the dredge-up events in low mass stars evolving along the Red Giant Branch: in the stellar case the convection acts as dredge and bring to the surface the product of nuclear burning altering the chemical composition of the star. In the planetary case, instead, it is the shock wave that acts as a dredge. The results of such a larger impact can alter the chemical composition of the planet for longer time-scales than smaller events like SL9. A repeated series of collisions of large impactor (i.e. with size > 10 km) can therefore permanently alter the chemical structure of the atmosphere of an extrasolar planet.

## 6. Toward the investigation of exoplanetary atmospheres in NIR/MIR

In view of the amounts of data that will be provided by future EChO-like space missions devoted to the systematic investigation of the atmospheric composition of extrasolar planets, in the next years it will be necessary to focus part of our attention on the strategies to link the atmospheric composition of an extrasolar planet to its past evolution and its formation region. As an example, Fig. 8 illustrates the steps that need to be undertaken to unveil the sources and sinks of the molecular species detected in the exoplanetary atmosphere of a giant planet (assuming it is the only one present in the planetary system). First, we would need to be able to assess whether the detected molecular species can form in situ due to the thermodynamical conditions in the atmosphere of the planet: if not, what we need to assess is whether they are stable or transient and, in the latter case, how long they can reside in the atmosphere before being destroyed or reprocessed (the “atmospheric” line of investigation identified by the orange arrows in Fig. 8). If the species are exogenous and transient, we need to assess whether secular processes like the impacts of cometary or asteroidal bodies have a reasonable chance of producing the observed abundances based on plausible fluxes and size frequency distributions of the impactors (the "secular contamination" line of investigation identified by the green arrows in Fig. 8). If the species are long-lived and/or secular effects cannot account for the observed abundances, then we need to reconstruct the composition of the gas in the circumstellar disk and identify if the regions that could represent the sources of said species are compatible with plausible dynamical histories of the giant planet (the "primordial accretion" line of investigation identified by the light blue arrows in Fig. 8).

As the previous discussion highlights, our present understanding of the formation and evolution of planetary systems, both around the Sun and other stars, allows us to have a more or less clear picture (from a qualitative point of view) of what are the processes and factors that can affect the composition of the atmospheres of giant planets. The same is not necessarily true, however, for the outcomes of said processes (especially from a quantitative point of view) and, consequently, for our capability to constrain the past histories of the planetary systems from the observations of the composition of planetary atmospheres. As schematically illustrated by Fig. 9 for what it concerns the problem of the migration of giant planets, we can track the past dynamical evolution of planetary systems in those cases where the planets did not migrate significantly (e.g. they accreted the gas while the disk was dissipating or got locked in resonant configurations that prevented migration) or where a single giant planet was present and its migration was induced solely by its interactions with the gas in the circumstellar disk (as assumed implicitly in the example of Fig. 8). More complex cases, like those where the final orbital configuration of the planetary system is created by the Jumping Jupiters mechanism, are much more difficult (if not impossible) to disentangle due to the intrinsic degeneracy of the problem. In the coming years, therefore, we will also need to devote several efforts to improve our comprehension of the atmospheric evolution of exoplanets and of how the past evolution of the extrasolar systems could have shaped them, in order to be prepared for the challenges that the new data will present.

## *Acknowledgments*

D.T. would like to thanks Alberto Adriani, Francesca Altieri, Maria Teresa Capria, Davide Grassi and Roberto Peron for the helpful discussions. D.T. and M.B. acknowledge the financial contribution from agreement no. I/022/12/0 between the Italian Space Agency (ASI) and the Italian National Institute for Astrophysics (INAF).

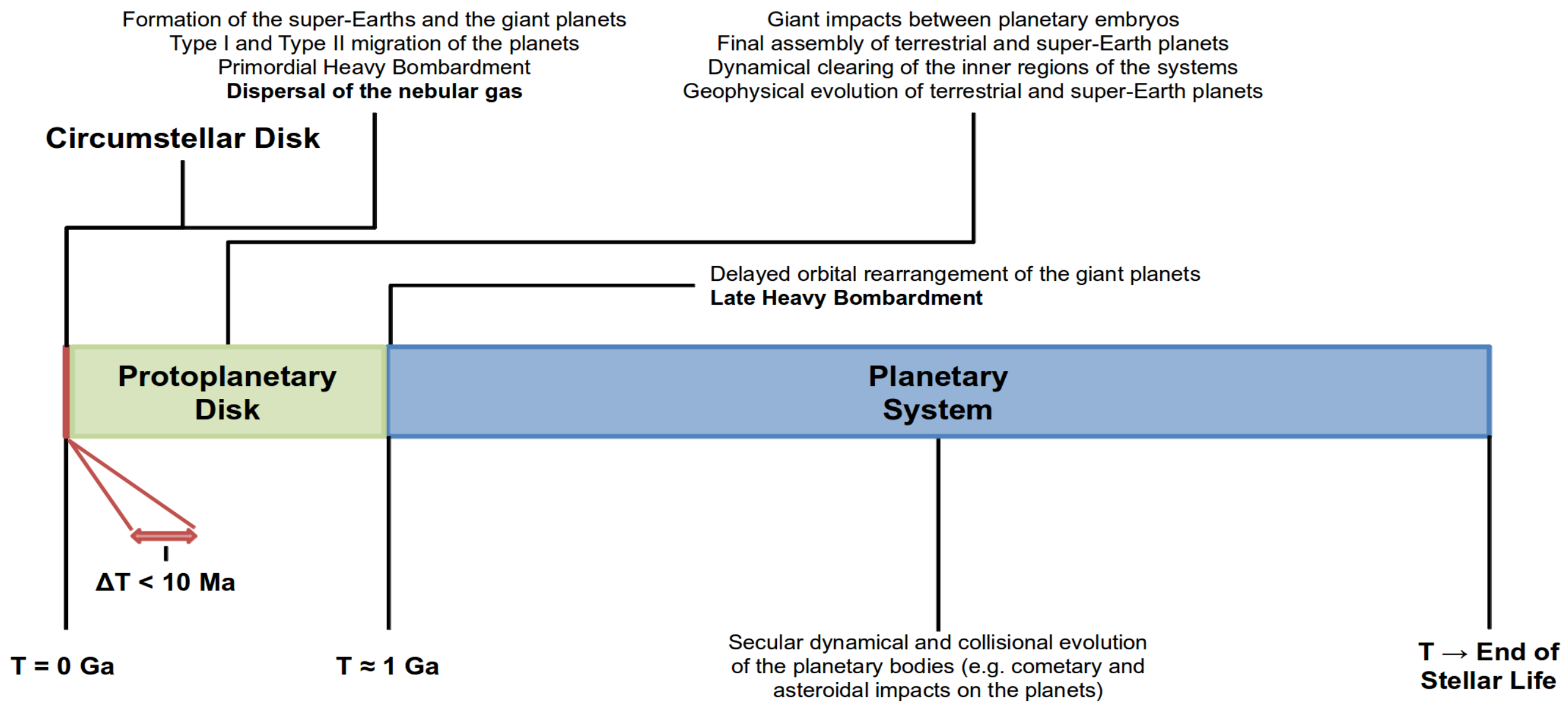

*Figure 1: timeline of the history of a generic planetary system following the division in three phases proposed by Coradini et al. (2011). The events marking the transition between the different phases are in bold characters.*

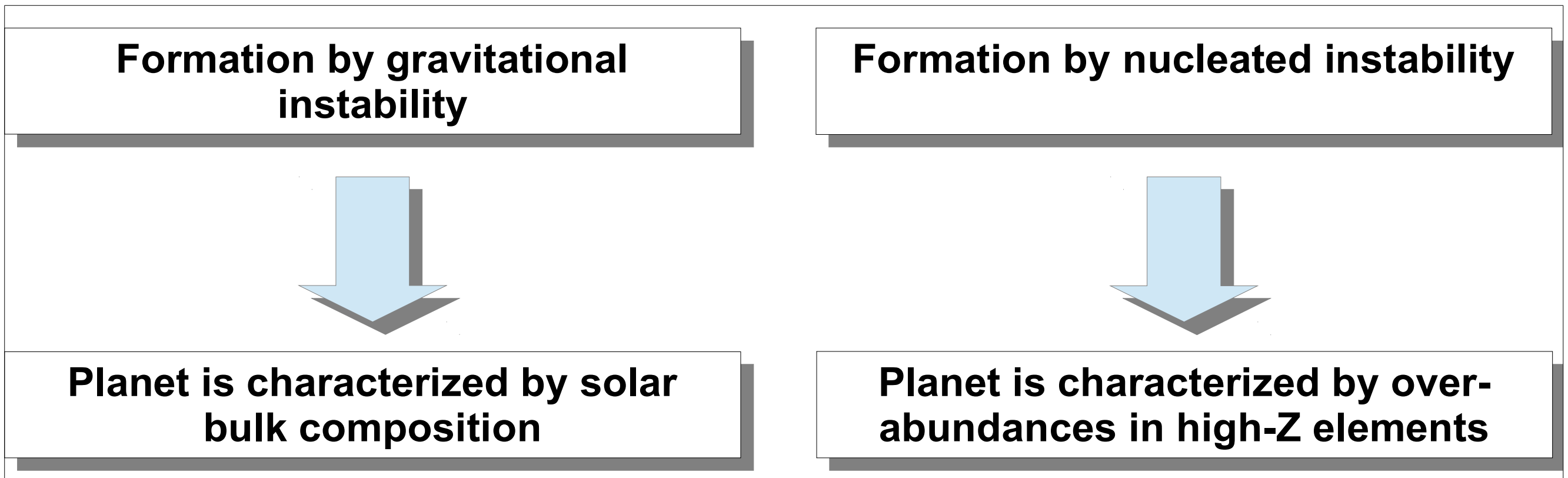

*Figure 2: Post-formation composition of a giant planet based on its formation process.*

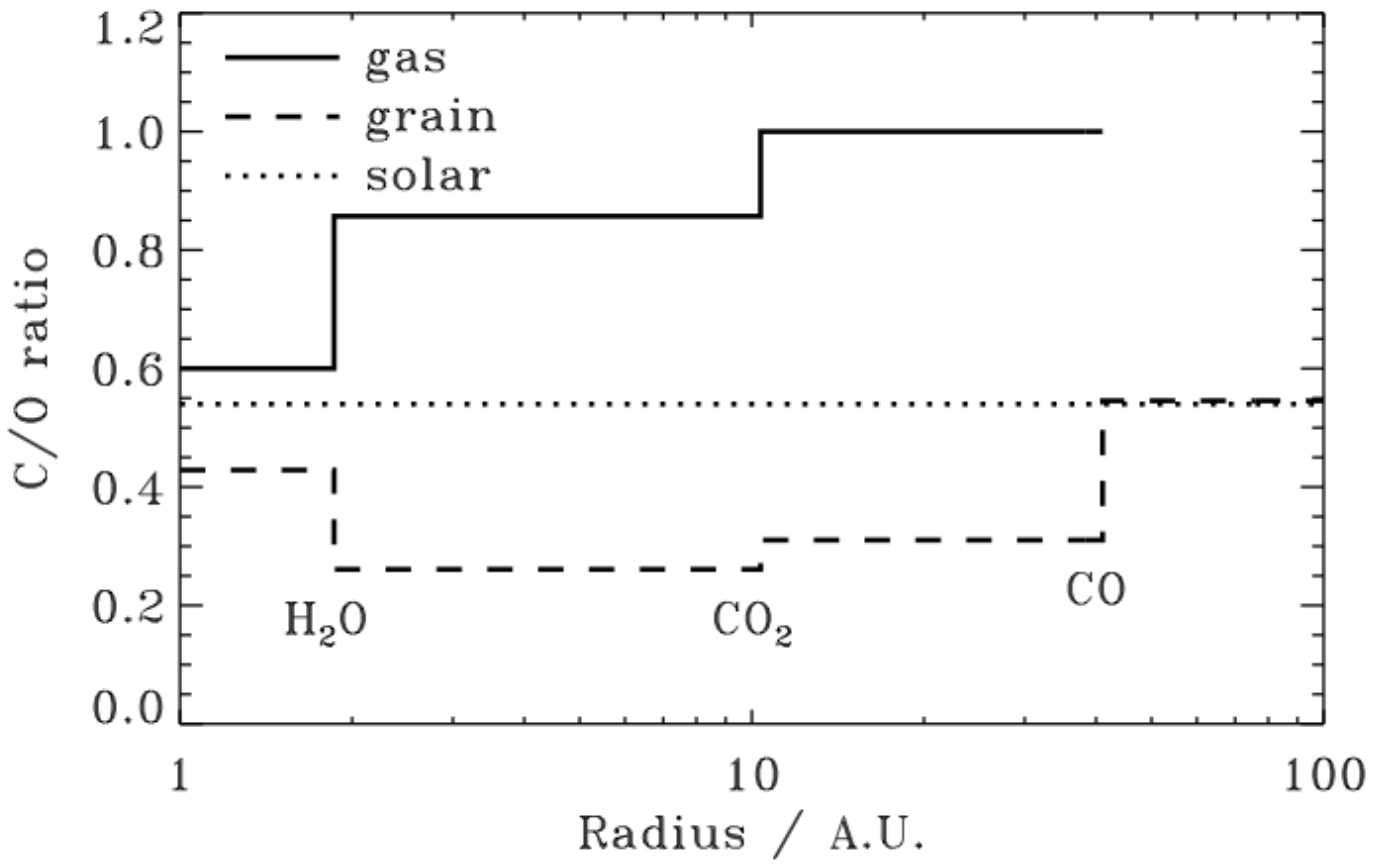

*Figure 3: Locations of the ice-lines and their influence on the C/O ratios for the gas and solids (adapted from Oberg et al. 2011).*

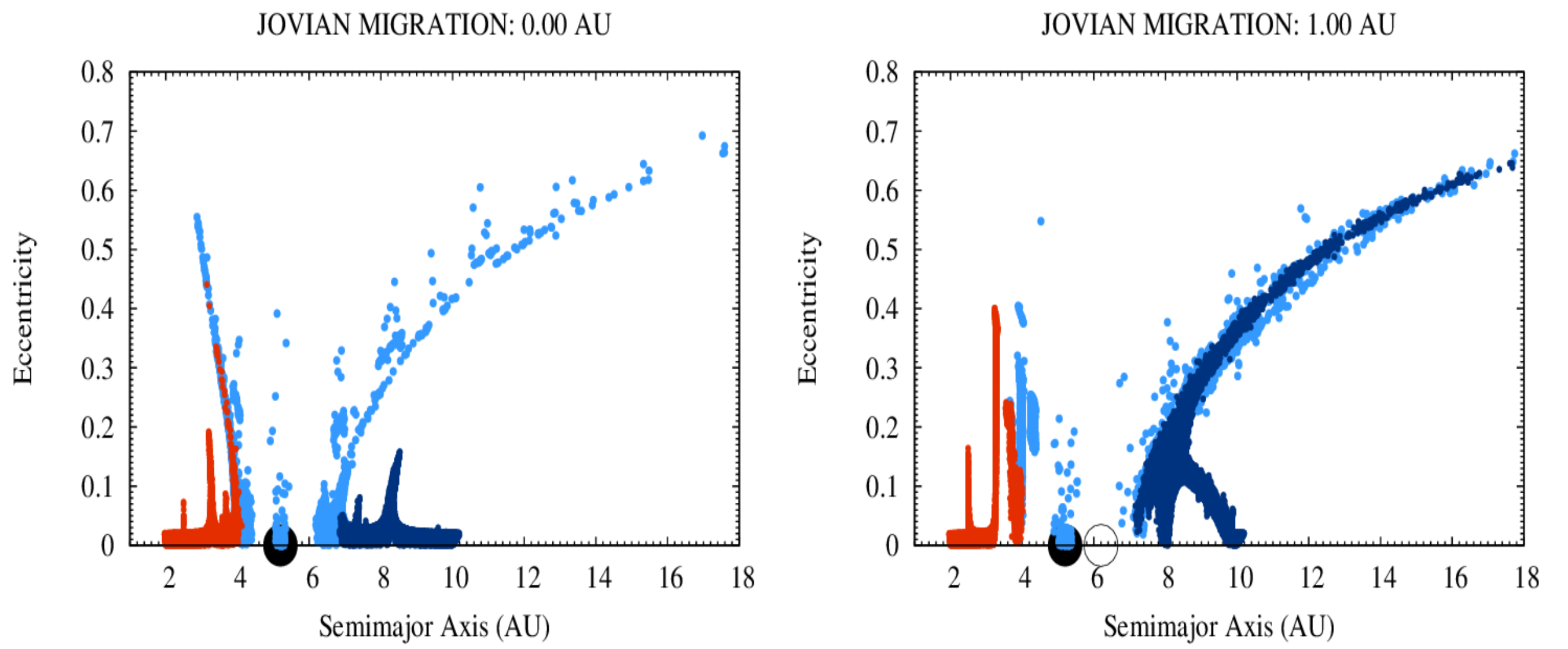

*Figure 4: orbital distribution of the Solar Nebula $2x10^5$ years after the beginning of the accretion of the nebular gas by Jupiter in the simulations performed by Turrini et al. (2011). Planetesimals that formed between 2 au and 4 au are indicated in red, those that formed between 4 au and 7 au in light blue and those that between 7 au and 10 au in dark blue. The open circle is the position of Jupiter at the beginning of the gas accretion phase, the filled one is the position of Jupiter once fully formed.*

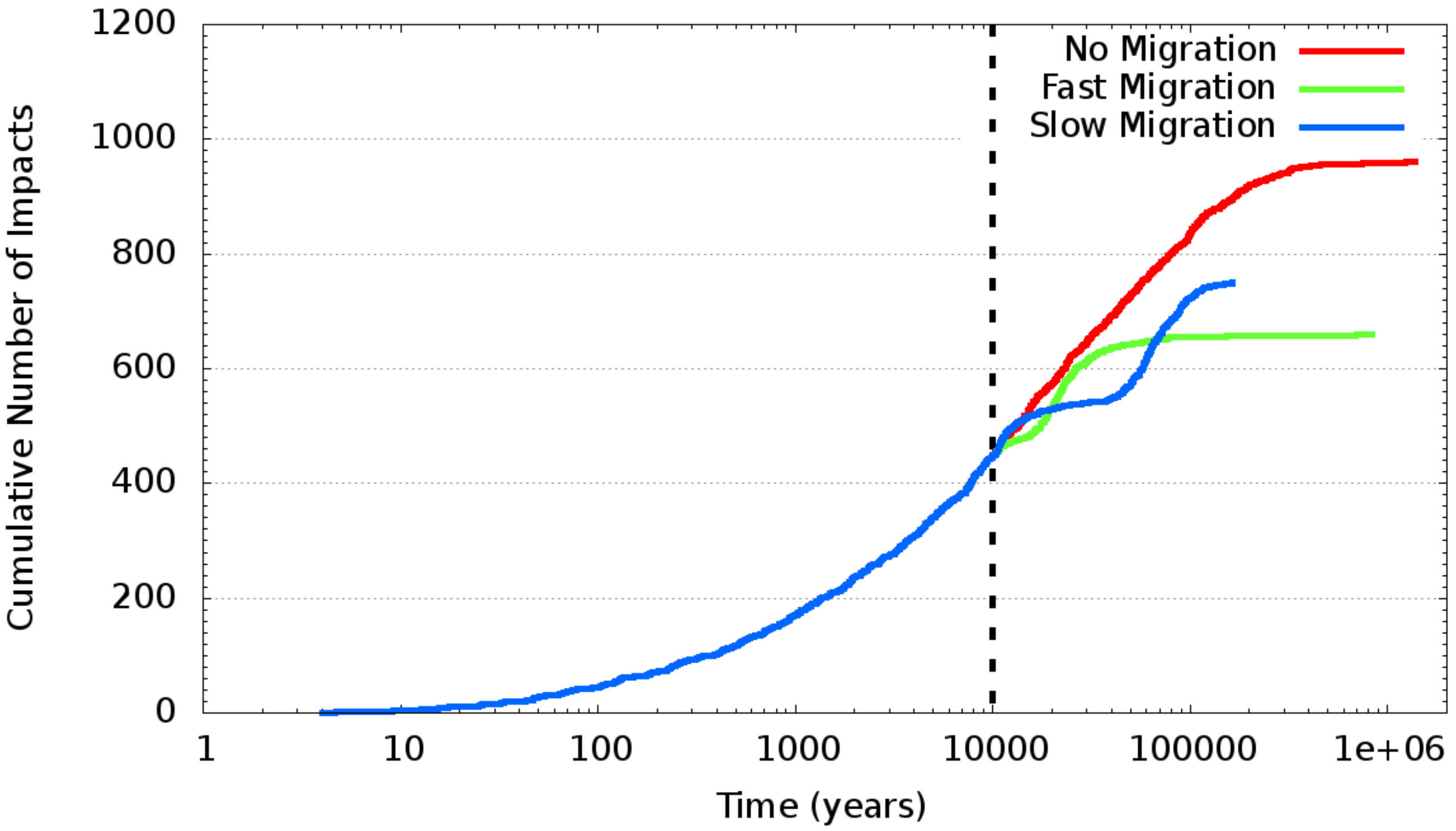

*Figure 5: Number of impacts on the giant planet as a function of time in the three scenarios considered. The vertical dashed line marks the beginning of the planetary migration in those scenario where it is present. The initial number of massless particles is 20000.*

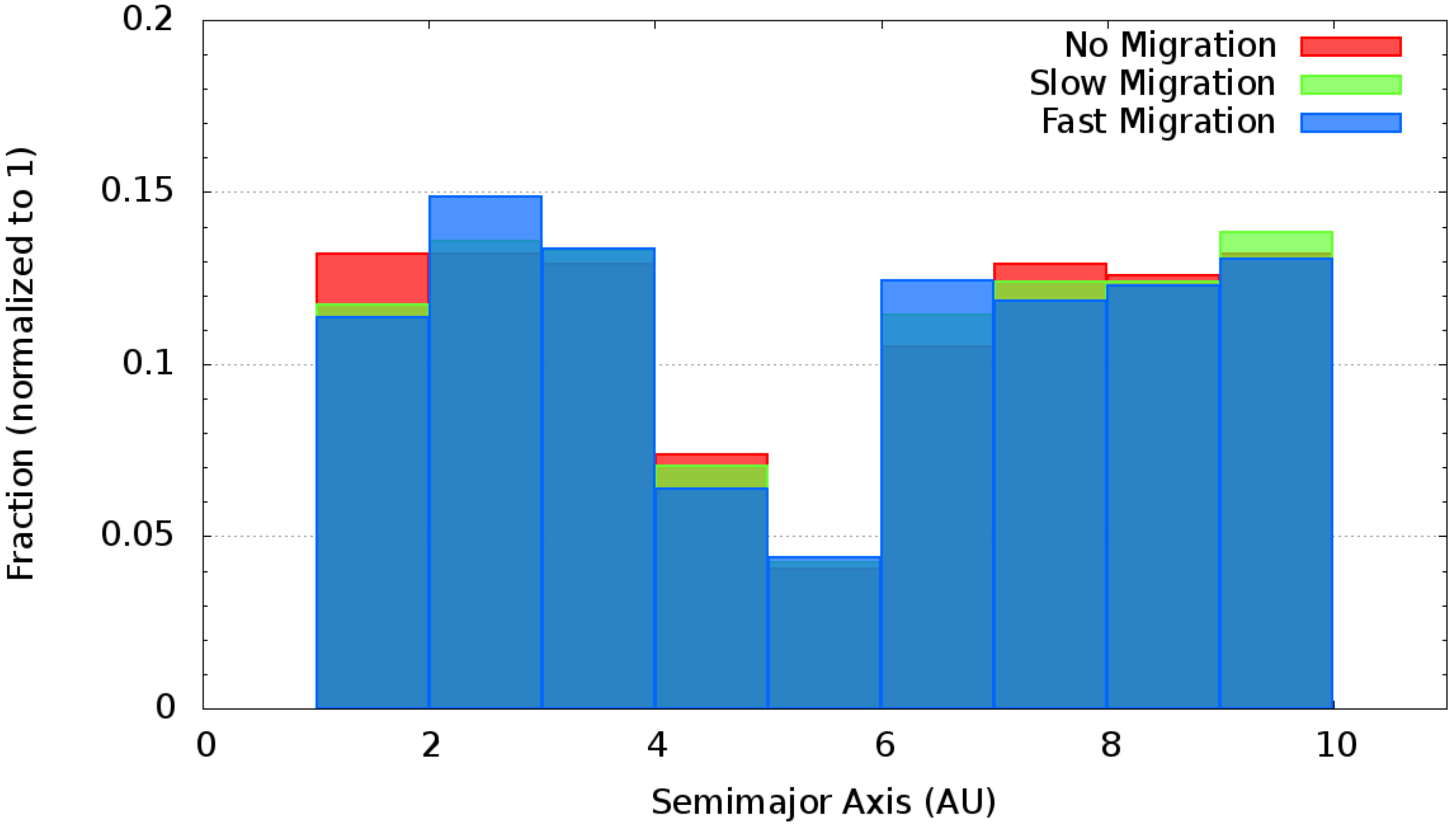

*Figure 6: Fractions of the capture particles coming from the different regions of the disk of massless particles (red for the no migration scenario, green and blue for the fast and slow migration scenarios respectively).*

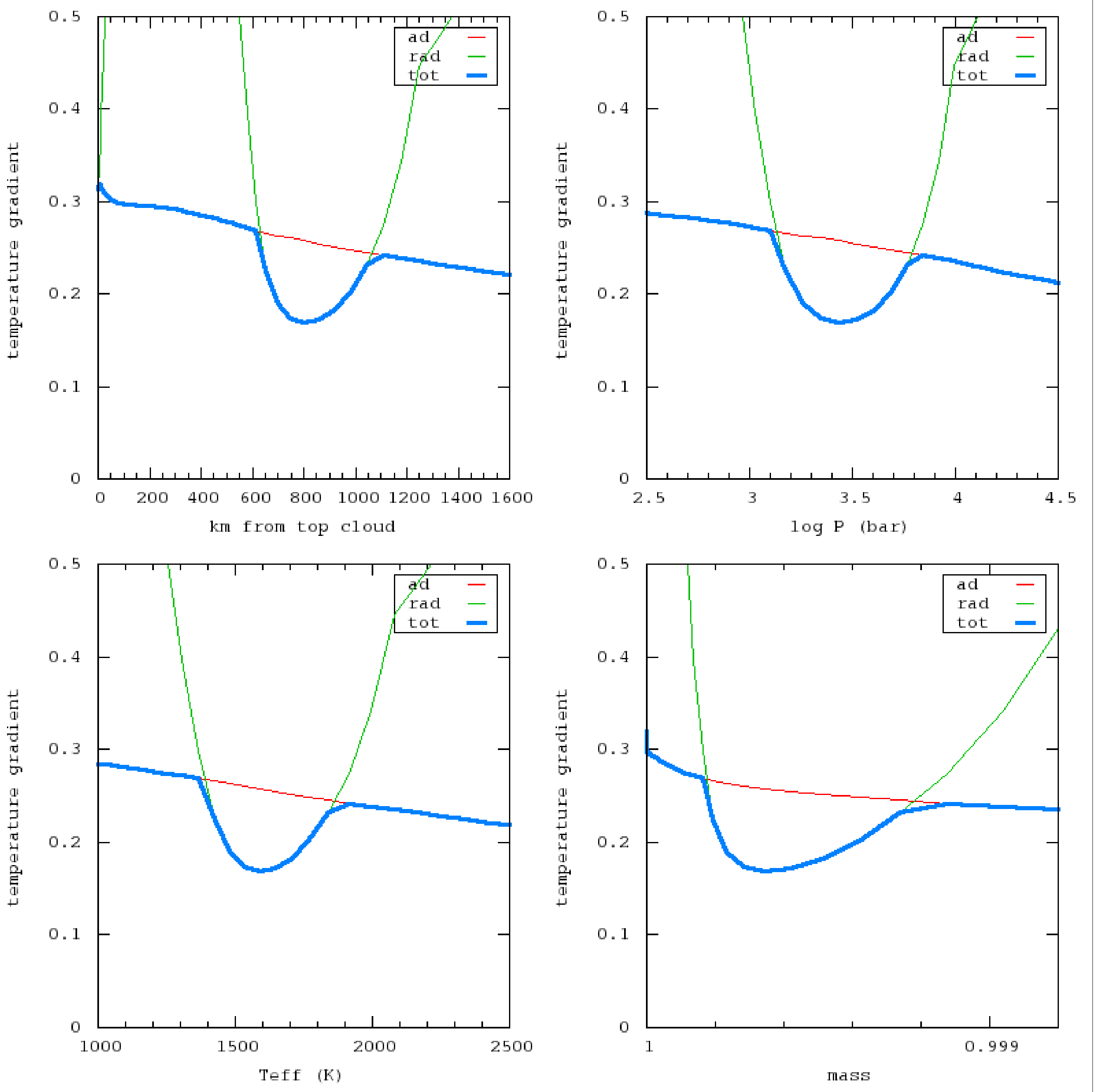

*Figure 7: Temperature gradients inside Jupiter from the models of Guillot et al. (1995). The gradients are: in red the adiabatic, in green the radiative, and in blue the total gradient. Each panel presents the gradients as function of different physical variables.*

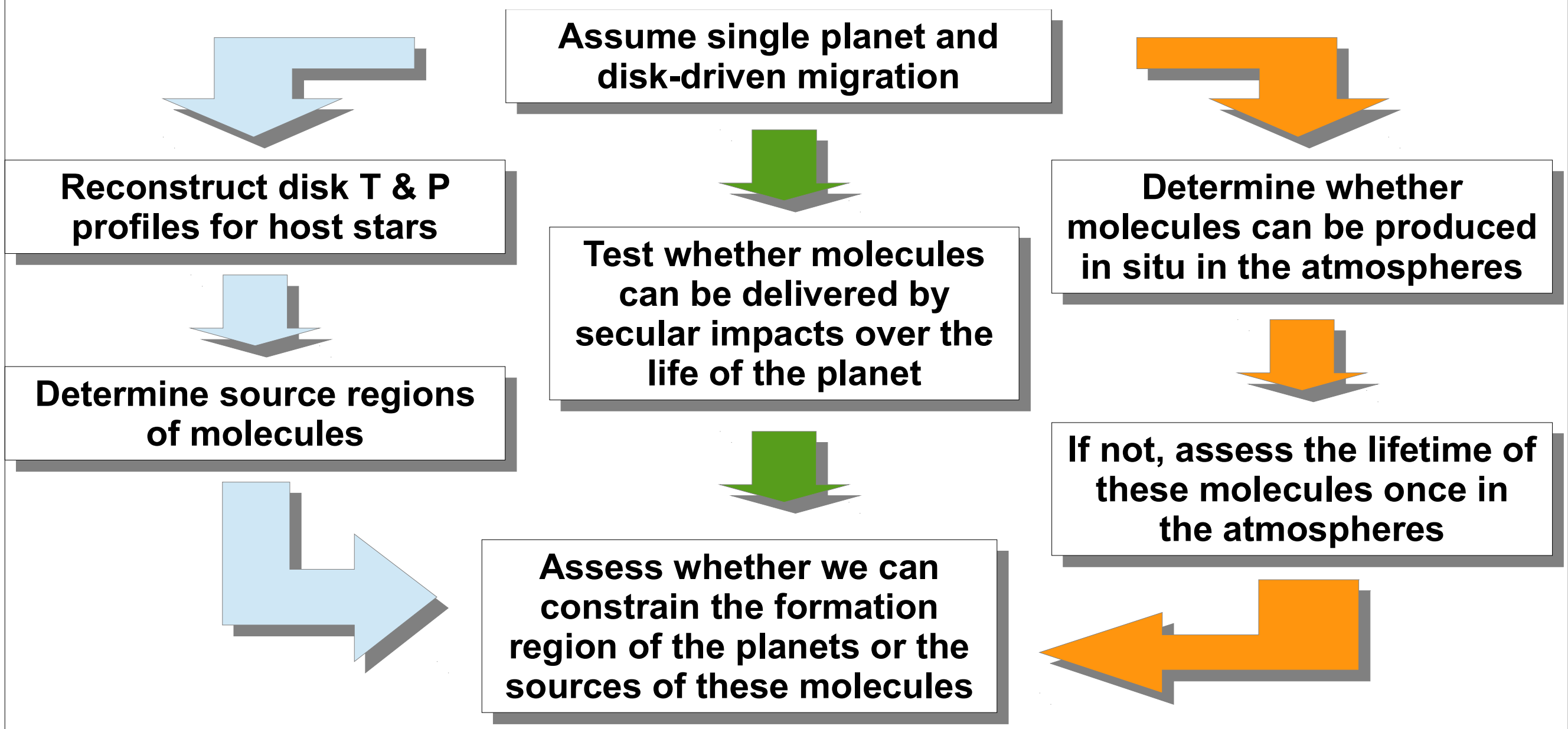

*Figure 8: A possible schematic approach to interpret the origins of molecular species in exoplanetary atmospheres.*

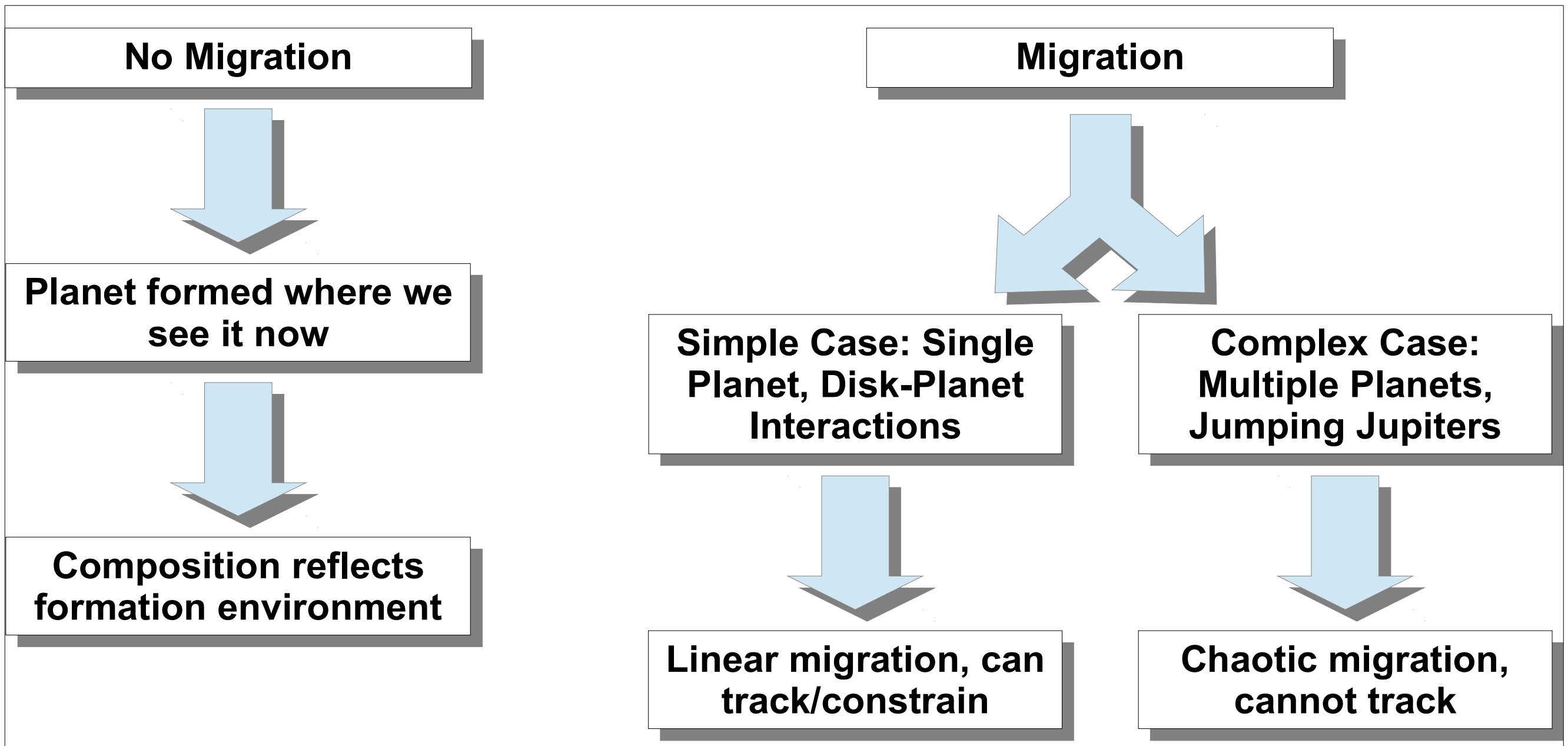

*Figure 9: Schematic representation of the possible dynamical paths of planetary systems and of how they affect our capability to identify their signatures in the atmospheric composition of the exoplanets.*